# Atmospheric neutrinos with the first detection units of KM3NeT/ARCA

A. Sinopoulou,[a,b,1] R. Coniglione,[c] R. Muller,[d] and E. Tzamariudaki,[a] on behalf of the KM3NeT Collaboration

[a] *NCSR "Demokritos", Institute of Nuclear and Particle Physics, 15310 Ag. Paraskevi, Athens, Greece*

[b] *National Technical University of Athens, School of Applied Mathematical and Physical Sciences, Zografou Campus, 9, Iroon Polytechniou str, 15780 Zografou, Athens, Greece*

[c] *INFN, Laboratori Nazionali del Sud, Via S. Sofia 62, Catania, 95123 Italy*

[d] *Nikhef Institute for Subatomic Physics, Science Park 105, 1098 XG Amsterdam, The Netherlands*

 *E-mail:* sinopoulou@inp.demokritos.gr

ABSTRACT: The KM3NeT Collaboration is constructing two deep-sea Cherenkov detectors in the Mediterranean Sea. The ARCA detector aims at TeV-PeV neutrino astronomy, while the ORCA detector is optimised for atmospheric neutrino oscillation studies at energies of a few GeV. In this contribution, an analysis of the data collected with the first deployed detection units of the ARCA detector is presented. A high-purity sample of atmospheric neutrinos is selected demonstrating the capability of the ARCA detector.

---

[1] Corresponding author.

# Contents

## 1. Introduction

KM3NeT is a research infrastructure that will host two neutrino detectors: ARCA (Astroparticle Research with Cosmics in the Abyss), located at a depth of 3500 m offshore Capo Passero (Italy), and ORCA (Oscillation Research with Cosmic in the Abyss), located at a depth of 2450 m offshore Toulon (France). Their location in the Mediterranean Sea offers an optimal window for the observation of the Southern sky, where the Galactic Centre and most of the Galactic Plane are located. The KM3NeT/ARCA detector, optimised for the neutrino energy range from few TeV to 100 PeV, focuses on high-energy neutrino astrophysics. The main detector component is the digital optical module (DOM), a pressure-resistant glass sphere housing 31 3-inch photomultiplier tubes (PMTs) and their associated electronics. The lower hemisphere of each DOM contains 19 PMTs, therefore downward looking, whereas the other 12 PMTs look upwards. The DOMs are arranged in string-like structures, called detection units (DUs), anchored to the seabed, and held vertically by the buoyancy of the DOMs as well as a dedicated buoy at the top. The vertical spacing between the DOMs along a DU is 36 m and the horizontal spacing between the DUs is ∼ 90 m for ARCA. In its final configuration, ARCA will consist of 230 DUs instrumenting a volume of ∼ 1 $\mathbf{km^3}$ of seawater. In this contribution, the very first data from the KM3NeT/ARCA configuration will be presented.

### 1.1 Different detector configurations – data taking periods

The first KM3NeT/ARCA DU was deployed in December 2015. Out of the two additional DUs deployed in May 2016, only one was operational. Therefore, data were collected with two DUs (ARCA2 period). After the commissioning and calibration of ARCA2, a stable and reliable detector performance was achieved. Different analyses have been performed in the past concerning ARCA2 as shown in [1],[2]. Due to consecutive upgrades of the seabed network infrastructure, ARCA detector operation was on hold between April 2017 and January 2019, after which, data taking continued with only one operating DU (ARCA1 period). In mid-April 2021, five more DUs were deployed at the ARCA site and a new data taking period started operating with 6 DUs (ARCA6 period).

The data taking period for ARCA2 in operation has an effective livetime of ~ 53 days. The total livetime of the reconstructed data for the second period analysed, ARCA1, is ~ 207 days. The atmospheric neutrino candidates presented concern the combined ARCA2 and ARCA1 data taking periods. To demonstrate the performance of the current ARCA configuration, data from the first weeks of operation with 6 DUs have been reconstructed. The total livetime of ARCA6 data corresponds to ~ 19 days.

## 2. Analysis

The main goal of ARCA will be the search for astrophysical neutrino sources and the detection of the diffuse astrophysical neutrino flux. For detectors focusing on neutrino astronomy, the background comes mainly from the contribution of atmospheric muons and neutrinos; therefore, the suppression of these contributions, as much as possible, is crucial for cosmic neutrino searches. Currently ARCA is at the construction phase therefore, the instrumented volume is still too small for significant studies in terms of neutrino astronomy. The purpose of this analysis is to demonstrate that with the already deployed configurations of ARCA it is possible to reject the atmospheric muon background and detect atmospheric neutrino candidates. This task is quite challenging for the first ARCA configurations with a low number of DUs, since the final ARCA configuration has been optimised for the detection of high energy neutrinos.

### 2.1 Event Selection

As already reported, the main physical background for identifying neutrino candidates comes from the atmospheric muon contribution. Atmospheric muons, produced in CR interactions in the atmosphere, reach the detector in a downward direction (from the upper to the down-most part of the detector) whereas those coming upwards through the Earth interact and are absorbed before arriving to the detector. Sometimes though, atmospheric muon events are misreconstructed as upward going, particularly when the detector volume is very limited. This analysis aims to reject those misreconstructed atmospheric muon events and constitutes a primary step for all the physics analyses in such experiments.

Detailed comparisons between data and Monte Carlo (MC) events were carried out in order to assess the detector performance. A MC sample of atmospheric muons with similar livetime to the one of the data (run-by-run generation) has been used. The distribution of the cosine of the reconstructed zenith angle is shown in Fig. 1 (left) for all the reconstructed data events (black), atmospheric muon simulated events (blue) and atmospheric neutrino simulated events (red) combined for ARCA1 and ARCA2 detector configurations. As expected, the contribution of the atmospheric muons dominates the observed rate. Atmospheric muon MC events misreconstructed as upgoing are populating the $\cos(\theta_{\text{zenith}}) < 0$ region. Since atmospheric neutrinos are isotropically distributed, the main goal is to differentiate them from the observed misreconstructed atmospheric muons. In order to do that, the first step is focused on studying and applying requirements to reject poorly reconstructed events. As a second step, dedicated requirements and a cut on the reconstructed zenith angle are applied in order to isolate atmospheric neutrino candidate events. In this sense, a variable indicating the quality of the track fit used for the event selection is the ratio of the number of PMTs having signal-like hits (hits with a small-time residual between the observed and the expected arrival time to the PMT assuming Cherenkov photon hypothesis) over all PMTs used for the track reconstruction. In Fig. 1 (right), the distribution of this ratio is shown for upgoing events combining the two detector configurations for the data (black), atmospheric muon MC (blue) and atmospheric neutrino MC (red) in an intermediate step of the analysis. One can notice that poorly reconstructed atmospheric muon events have less PMTs with signal-like hits compared to atmospheric neutrino events which are correctly reconstructed as upward going.

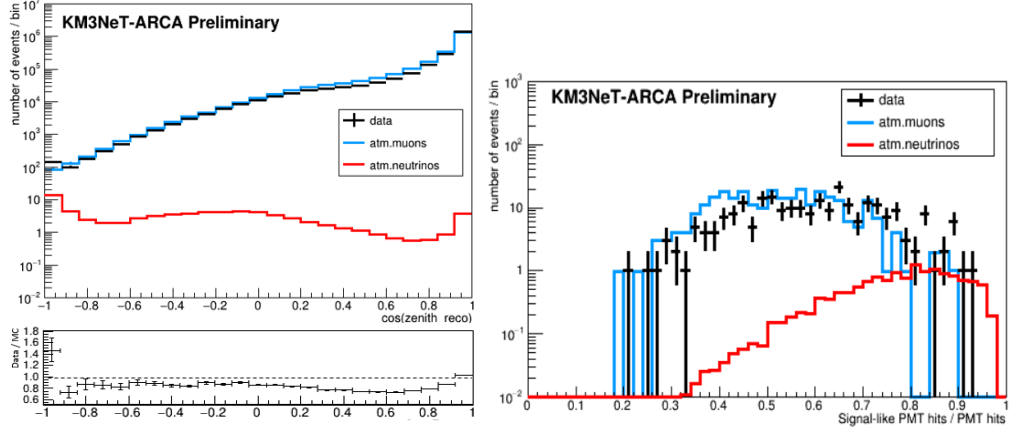

**Figure 1.** (Left) Distribution of the cosine of the zenith angle for all reconstructed events. (Right) Ratio of the number of PMTs having signal-like hits over all PMTs in an intermediate level of the analysis. The data shown reflect to ~1% of the total reconstructed data.

## 2.2 Atmospheric neutrino candidates

The distribution of the cosine of the reconstructed zenith angle is shown in Fig. 2 for all the events surviving the selection criteria. ARCA1 and ARCA2 operation periods have been combined for data (black), atmospheric muon MC (blue) and atmospheric neutrino MC (red). A powerful rejection of atmospheric muons misreconstructed as upgoing has been achieved as only one of these events survives the final selection. A total number of 15 neutrino candidates have been found with $\cos(\theta_{\text{zenith}}) < 0$, while 8.1 are expected from the atmospheric neutrino MC sample in ~ 260 days of total livetime.

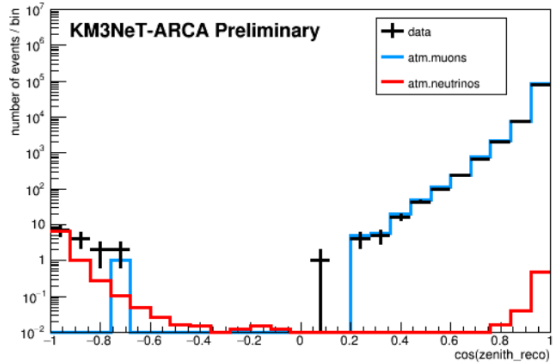

**Figure 2.** Distribution of the cosine of the zenith angle for events surviving the neutrino selection criteria.

## 3. Current ARCA configuration – ARCA6

ARCA is currently operating with 6 DUs since the end of April 2021. The data analysed in this contribution have an effective livetime of ~ 19 days. Due to the limited livetime, loose quality selection requirements were investigated and applied in order to reject misreconstructed atmospheric muon events. The distribution of the cosine of the reconstructed zenith angle is shown in Fig. 3 (left) for all the reconstructed data events (black), atmospheric muon simulated events (blue) and atmospheric neutrino simulated events (red) for the ARCA6 detector configuration. A reasonable data and MC agreement is observed.

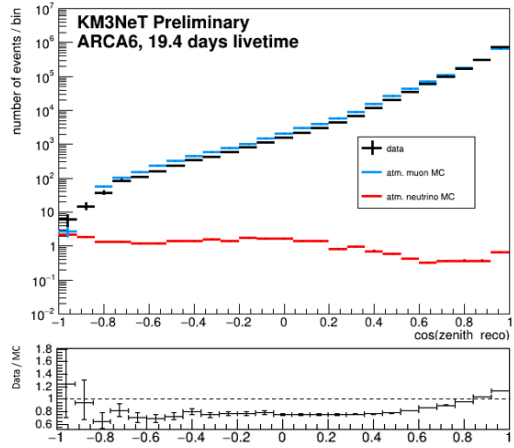

**Figure 3.** Distribution of the cosine of the zenith angle for all reconstructed events.

An excellent angular resolution of less than 0.1º for $E > 100$ TeV for the full ARCA detector is expected, due to the innovative multi-PMT DOMs of KM3NeT and the large scattering length of deep-sea water. For the current ARCA configuration, the angular resolution is shown after the quality selection requirements in Fig. 4 (left) as a function of the true (MC) neutrino energy. The median of the angular resolution is 0.75º.

In order to investigate the prospects of the current configuration, a comparison of the detector performance in terms of the effective area is shown for ARCA6, the current ORCA configuration (ORCA6) and ANTARES in Fig. 4 (right). For all the reconstructed upgoing neutrino MC events in the low energy range ($E < 1$ TeV), ORCA6 has a slighter higher effective area than ARCA6. However, for higher energies ($E > 100$ TeV), for which ARCA is optimised, ARCA6 is expected to perform significantly better than ORCA6 and ANTARES.

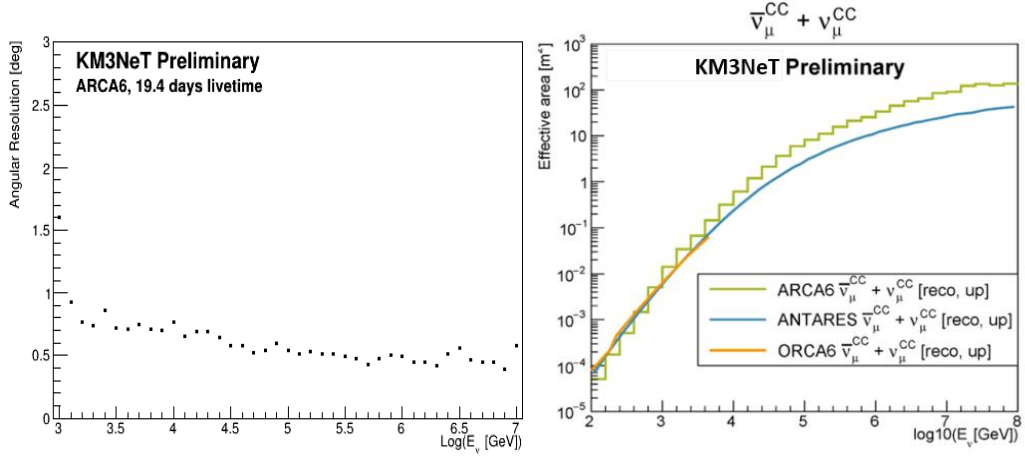

**Figure 4.** (Left) Angular resolution for events surviving the selection criteria as a function of the true neutrino energy. (Right) Effective area for all neutrino MC events reconstructed as upgoing for ARCA6 (green), ANTARES (blue) and ORCA6 (orange).

## 4. Summary

For the past and current ARCA detector configurations, the search for atmospheric neutrino candidates is more challenging than for ORCA due to the higher neutrino energy detection threshold, the very limited detector volume and the small effective livetime of ARCA. Within a period of ∼ 260 days of effective livetime, with one (ARCA1) and two (ARCA2) operational DUs, 15 neutrino candidates have been observed, with one atmospheric muon expected from the MC surviving the final selection criteria.

With the current ARCA configuration of 6 operational DUs (ARCA6) and an effective livetime of 19 days, a reasonable data and MC agreement is observed for all the reconstructed events and for events surviving loose selection requirements. The median of the angular resolution after the event selection is $0.75^0$. In terms of performance, for the low energy range ($E < 1$ TeV) the current ORCA configuration (ORCA6) is expected to have slightly higher effective area than ARCA6. In the energy range of interest for ARCA ($E > 100$ TeV), ARCA6 is expected to perform significantly better than ORCA6 and ANTARES.

The aim of this analysis has been to show that with a 1-3% of the full instrumented volume of the ARCA detector, it is possible to detect atmospheric neutrinos and to achieve a powerful reduction of the atmospheric muon contribution. Furthermore, a good data/MC agreement verifies the KM3NeT technology, the detector understanding and detector calibration demonstrating the capability of the future KM3NeT detectors.

## Acknowledgments

Anna Sinopoulou acknowledges the support of the Hellenic Foundation for Research and Innovation (H.F.R.I.) under the HFRI PhD Fellowship grant (Fellowship Number (1076)).

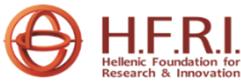